\documentclass[english]{article}
\usepackage[T1]{fontenc}
\usepackage[latin9]{inputenc}
\usepackage{amsmath}
\usepackage{amssymb}
\usepackage{graphicx}

\makeatletter

\providecommand{\tabularnewline}{\\}

\usepackage{INTERSPEECH2020}
\usepackage[bookmarks=false,pdftex,pdfpagemode=None,pdfstartview=FitH,pdfview=FitH,pdftitle={A Perceptually-Motivated Approach for Low-Complexity, Real-Time Enhancement of Fullband Speech},pdfauthor={Jean-Marc Valin, Umut Isik, Neerad Phansalkar, Ritwik Giri, Karim Helwani, Arvindh Krishnaswamy}]{hyperref}

\makeatother

\usepackage{babel}
\begin{document}
\name{Jean-Marc Valin, Umut Isik, Neerad Phansalkar, Ritwik Giri,\\ Karim Helwani, Arvindh Krishnaswamy}
\address{Amazon Web Services}
\email{\{jmvalin, umutisik, neeradp, ritwikg, helwk, arvindhk\}@amazon.com}

\title{A Perceptually-Motivated Approach for Low-Complexity, Real-Time Enhancement
of Fullband Speech}
\maketitle
\begin{abstract}
Over the past few years, speech enhancement methods based on deep
learning have greatly surpassed traditional methods based on spectral
subtraction and spectral estimation. Many of these new techniques
operate directly in the the short-time Fourier transform (STFT) domain,
resulting in a high computational complexity. In this work, we propose
PercepNet, an efficient approach that relies on human perception of
speech by focusing on the spectral envelope and on the periodicity
of the speech. We demonstrate high-quality, real-time enhancement
of fullband (48 kHz) speech with less than 5\% of a CPU core.
\end{abstract}

\section{Introduction}

Over the past few years, speech enhancement methods based on deep
learning have greatly surpassed traditional methods based on spectral
subtraction~\cite{boll1979suppression} and spectral estimation~\cite{ephraim1985speech}.
Many of these techniques operate directly on the short-time Fourier
transform (STFT), estimating either magnitudes~\cite{liu2014experiments,xu2015regression,tan2018convolutional}
or ideal ratio masks (IRM)~\cite{narayanan2013ideal,zhao2016dnn}.
This typically requires a large number of neurons and weights, resulting
in a high complexity. It also partly explains why many of those methods
are restricted to 8~or 16~kHz. The use of the STFT also brings up
a trade-off with the window length -- long windows can cause musical
noise and reverb-like effects, whereas short windows do not provide
sufficient frequency resolution for removing noise between pitch harmonics.
These problems can be mitigated by the use of complex ratio masks~\cite{williamson2015complex}
or time-domain processing~\cite{pascual2017segan,rethage2018wavenet,macartney2018improved},
at the cost of further increasing complexity.

We propose PercepNet, an efficient approach that relies heavily on
human perception of speech signals and improves on RNNoise~\cite{valin2018rnnoise}.
More precisely, we rely on the perception of audio in critical bands
(Section~\ref{sec:signal-model}) and on the perception of tones
and noise (Section~\ref{sec:pitch-filtering}) with a new acausal
comb filter. The deep neural network (DNN) model we use is trained
using perceptual criteria (Section~\ref{sec:DNN-model}). We propose
a novel envelope postfilter (Section~\ref{sec:post-filtering}) that
further improves the enhanced signal. 

The PercepNet algorithm operates on 10\nobreakdash-ms frames with
40~ms of look-ahead and can enhance 48~kHz speech in real time using
just 4.1\% of an x86 CPU core. We show that its quality significantly
exceeds that of RNNoise (Section~\ref{sec:results}).

\section{Signal Model}

\label{sec:signal-model}

Let $x\left(n\right)$ be a clean speech signal, the signal captured
by a hands-free microphone in a noisy room is given by
\begin{equation}
y\left(n\right)=x\left(n\right)\star h\left(n\right)+\eta\left(n\right)\,,\label{eq:additive-noise}
\end{equation}
where $\eta\left(n\right)$ is the additive noise from the room, $h\left(n\right)$
is the impulse response from the talker to the microphone, and $\star$~denotes
the convolution. Furthermore, the clean speech can be expressed as
$x\left(n\right)=p\left(n\right)+u\left(n\right)$, where $p\left(n\right)$
is a locally periodic component and $u\left(n\right)$ is a stochastic
component (here we consider transients such as stops as part of the
stochastic component). In this work, we attempt to compute an enhanced
signal $\hat{x}\left(n\right)=\hat{p}\left(n\right)+\hat{u}\left(n\right)$
which is as perceptually close to the clean speech $x\left(n\right)$
as possible. Separating the stochastic component $u\left(n\right)$
from the environmental noise $\eta\left(n\right)$ is a very hard
problem. Fortunately, we only need $\hat{u}\left(n\right)$ to \emph{sound}
like $u\left(n\right)$, which can be achieved by filtering the mixture
$u\left(n\right)\star h\left(n\right)+\eta\left(n\right)$ to have
the same spectral envelope as $u\left(n\right)$. Since $p\left(n\right)$
is periodic and the noise is assumed not to have strong periodicity,
$\hat{p}\left(n\right)$ should be easier to estimate. Again, we mostly
need $\hat{p}\left(n\right)$ to have the same spectral envelope and
the same period as $p\left(n\right)$.

We seek to construct an enhanced signal with the same 1) spectral
envelope, and 2) frequency-dependent periodic-to-stochastic ratio,
as the clean signal. For both these properties, we use a resolution
that matches human perception.

\begin{figure}
\centering{\includegraphics[width=1\columnwidth]{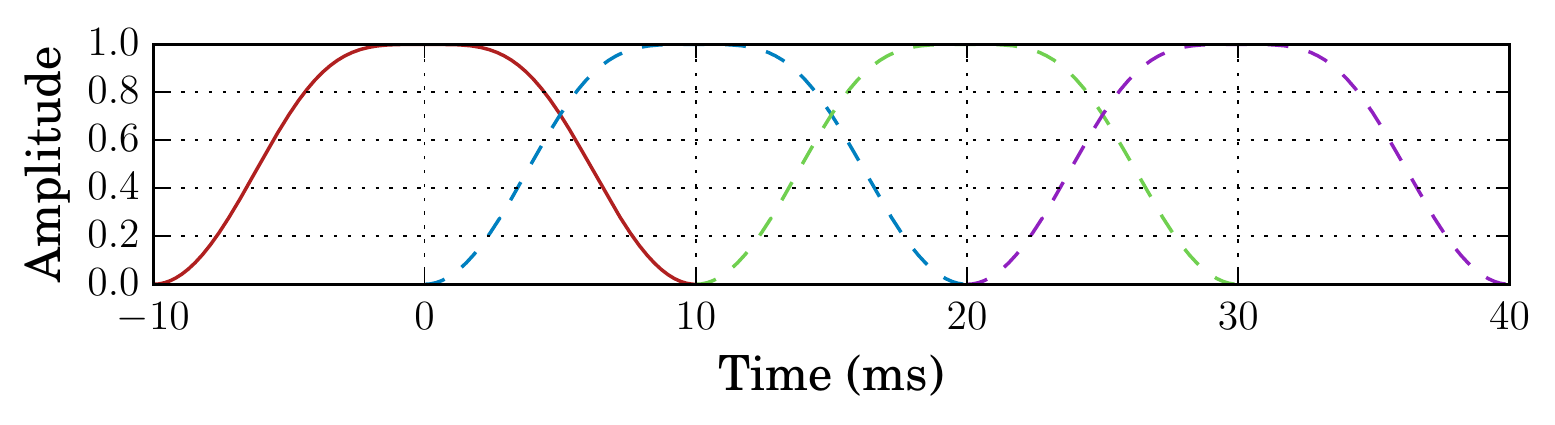}}\caption{The current window being synthesized is shown in solid red. We use
three windows of look-ahead (shown in dashed lines) such that samples
up to time $t=40\,ms$ are used to compute the audio output up to
$t=0$.\label{fig:windowing} }

\end{figure}

\begin{figure}
\centering{\includegraphics[width=1\columnwidth]{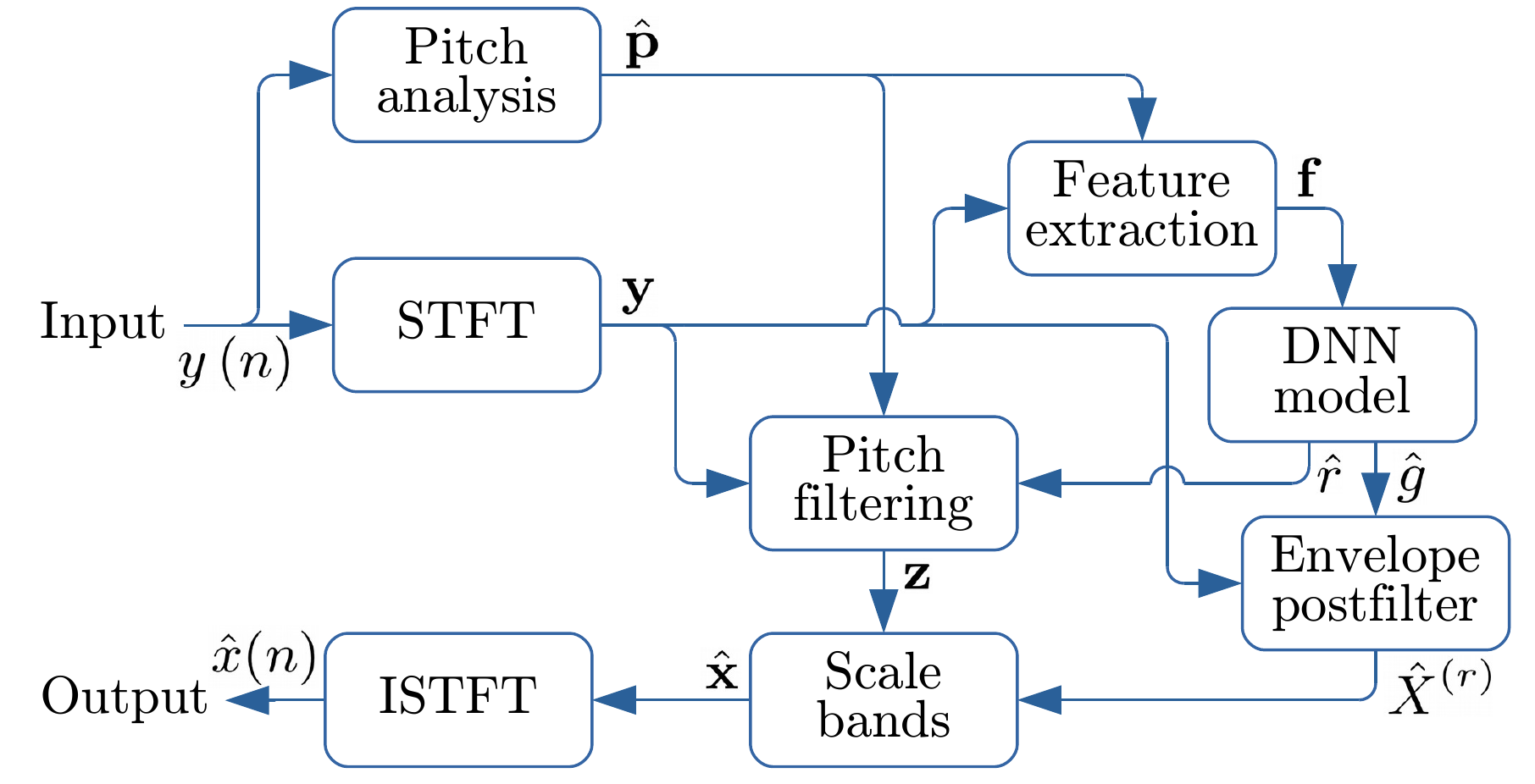}}\caption{Overview of the PercepNet algorithm.\label{fig:Overview-of-algorithm}}
\end{figure}

We use the short-time Fourier transform (STFT) with 20\nobreakdash-ms
windows and 50\%~overlap. We use the Vorbis window function~\cite{montgomery2004vorbis}
-- which satisfies the Princen-Bradley perfect reconstruction criterion~\cite{princen1986analysis}
-- for analysis and synthesis, as shown in Fig.~\ref{fig:windowing}.
An overview of the algorithm is shown in Fig.~\ref{fig:Overview-of-algorithm}.

\subsection{Bands}

The vast majority of noise signals have a wide bandwidth with a smooth
spectrum. Similarly, both the periodic and the stochastic components
of speech have a smooth spectral envelope. This allows us to represent
their envelope from 0~to 20~kHz using 34~bands, spaced according
to the human hearing equivalent rectangular bandwidth~(ERB)~\cite{moore2012introduction}.
To avoid bands with just one DFT bin, we impose a minimum band width
of 100~Hz.

For each band of the enhanced signal to be perceptually close to the
clean speech, both their total energy and their periodic content should
be the same. In this paper, we denote the complex-valued spectrum
of the signal $x\left(n\right)$ for band $b$ in frame $\ell$ as
$\mathbf{x}_{b}\left(\ell\right)$. We also denote the $L_{2}$-norm
of that band as $X_{b}\left(\ell\right)$.

\subsection{Gains}

From the magnitude of the noisy speech signal in band~$b$, we compute
the ideal ratio mask, i.e. the gain that needs to be applied to $\mathbf{y}_{b}$
such that it has the same energy as $\mathbf{x}_{b}\left(\ell\right)$:
\begin{equation}
g_{b}\left(\ell\right)=\frac{X_{b}\left(\ell\right)}{Y_{b}\left(\ell\right)}\,.\label{eq:magnitude-gain}
\end{equation}
In the case where the speech only has a stochastic component, applying
the gain $g_{b}\left(\ell\right)$ to the magnitude spectrum in band
$b$ should result in an enhanced signal that is almost indistinguishable
from the clean speech signal. On the other hand, when the speech is
perfectly periodic, applying the gain $g_{b}\left(\ell\right)$ results
in an enhanced signal that sounds \emph{rougher} than the clean speech;
even though the energy is the same, the enhanced signal is less harmonic
than the clean speech. In that case, the noise is particularly perceptible
due to the fact that tones have relatively little masking effect on
noise~\cite{gockel2003asymmetry}. In that situation, we use the
comb filter described in the next section to remove the noise between
the pitch harmonics and make the signal more periodic.

\section{Pitch Filtering}

\label{sec:pitch-filtering}

To reconstruct the harmonic properties of the clean speech, we use
comb filtering based on the pitch frequency. The comb filter can achieve
much finer frequency resolution than would otherwise be possible with
the STFT (50~Hz using 20\nobreakdash-ms frames). We estimate the
pitch period using a correlation-based method combined with a dynamic
programming search~\cite{RAPT}.

\subsection{Filter}

\begin{figure}
\vspace{-0.20em}\centering{\includegraphics[width=1\columnwidth]{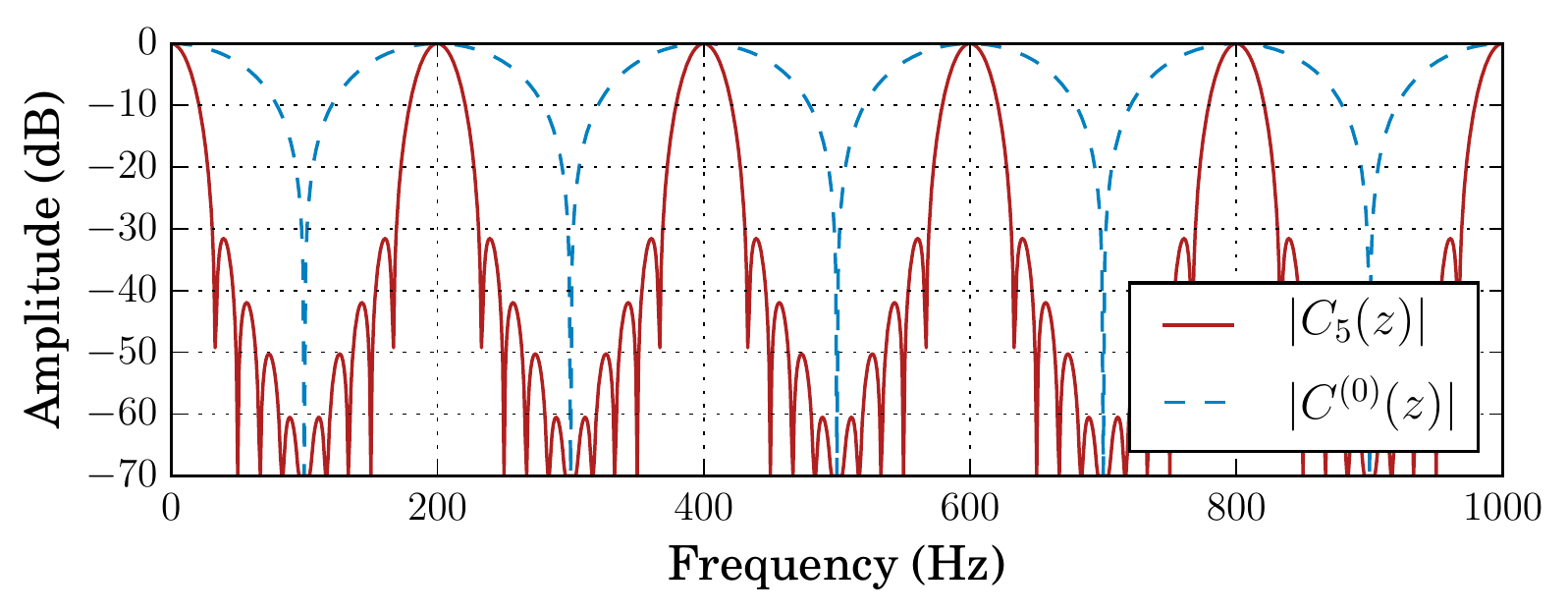}}\vspace{-0.5em}\caption{Frequency response of the proposed comb filter (red) vs the filter
used in~\cite{valin2018rnnoise} (blue) for a pitch of 200~Hz.\label{fig:Frequency-response-pitch}}
\end{figure}

For a voiced speech signal with period $T$, a simple comb filter
\begin{equation}
C^{(0)}(z)=\frac{1+z^{-T}}{2}\label{eq:simple-pitch}
\end{equation}
introduces zeros at regular interval between harmonics and attenuates
the noisy part of the signal by around 3~dB. This provided a small,
but noticeable quality improvement in~\cite{valin2018rnnoise}. In
this work, we extend the comb filtering to more than one period, including
non-causal taps using the following filter:
\begin{equation}
C_{M}(z)=\sum_{k=-M}^{M}w_{k}z^{-kT}\,,\label{eq:multi-period-filter}
\end{equation}
where $M$ is the number of periods on each side of the central tap
and $w_{k}$ is a window function satisfying $\sum_{k}w_{k}=1$. Using
$C_{M}\left(z\right)$, the noisy part of the signal is attenuated
by $\sigma_{w}^{2}=\sum_{k}w_{k}^{2}$ . Although a rectangular window
would minimize $\sigma_{w}^{2}$, we use a Hann window, which shapes
the remaining noise to be lower between harmonics. Due to the behavior
of tone masking~\cite{moore2012introduction}, this results in a
lower perceptual noise. For $M=5$, we have $\sigma_{w}=-9\,\mathrm{dB}$
and the full response is shown in Fig.~\ref{fig:Frequency-response-pitch}.
In practice, since the maximum look-ahead is bounded, we truncate
the window $w_{k}$ to values of $kT$ that are permitted. 

The filtering occurs in the time domain, with the output denoted $\hat{p}\left(n\right)$
since it approximates the ``perfect'' periodic component $p\left(n\right)$
from the clean speech. Its STFT is denoted $\hat{\mathbf{p}}_{b}\left(\ell\right)$.

\subsection{Filtering Strength}

\label{sub:filtering-strength}

The amount of comb filtering is important: not enough filtering results
in roughness, whereas too much results in a robotic voice. The strength
of the comb filtering in~\cite{valin2018rnnoise} is controlled by
a heuristic. In this work, we instead have the neural network learn
the strength that best preserves the ratio of periodic to stochastic
energy in each band. The equations below describe what that ideal
strength should be. Since they rely on properties of the clean speech,
they are only used at training time.

We define the pitch coherence $q_{x,b}\left(\ell\right)$ of the clean
signal as the cosine distance between the complex spectra of the signal
with its periodic component (both $\ell$ and $b$ are omitted for
clarity) 
\begin{equation}
q_{x}\triangleq\frac{\Re\left[\mathbf{p}^{\mathrm{H}}\mathbf{x}\right]}{\left\Vert \mathbf{p}\right\Vert \cdot\left\Vert \mathbf{x}\right\Vert }\,,\label{eq:pitch-coherence}
\end{equation}
where $\cdot^{\mathrm{H}}$ denotes the Hermitian transpose and $\Re\left[\cdot\right]$
denotes the real component. Similarly, we define $q_{y}$ as the pitch
coherence of the noisy signal. Since the ground truth $\mathbf{p}$
is not available, the coherence values need to be estimated. Considering
that the noise in $\hat{\mathbf{p}}$ is attenuated by a factor $\sigma_{w}^{2}$,
the pitch coherence of the estimated periodic signal $\hat{\mathbf{p}}$
itself can be approximated as
\begin{equation}
q_{\hat{p}}=\frac{q_{y}}{\sqrt{\left(1-\sigma_{w}^{2}\right)q_{y}^{2}+\sigma_{w}^{2}}}\,.\label{eq:pitch-enh-coherence}
\end{equation}

We define the pitch filtering strength $r\in\left[0,\,1\right]$,
where $r=0$ causes no filtering to occur and $r=1$ replaces the
signal with $\hat{\mathbf{p}}$. Let $\mathbf{z}=\left(1-r\right)\mathbf{y}+r\hat{\mathbf{p}}$
be a pitch-enhanced signal, we want the pitch coherence of $\mathbf{z}$
to match the clean signal:
\begin{equation}
q_{z}=\frac{\mathbf{p}\cdot\left(\left(1-r\right)\mathbf{y}+r\mathbf{\hat{\mathbf{p}}}\right)}{\left\Vert \mathbf{p}\right\Vert \cdot\left\Vert \left(1-r\right)\mathbf{y}+r\hat{\mathbf{p}}\right\Vert }=q_{x}\,.\label{eq:enhanced-coherence}
\end{equation}
Solving (\ref{eq:enhanced-coherence}) for $r$ results in
\begin{align}
r & =\frac{\alpha}{1+\alpha}\,,\label{eq:pitch-strength-norm}\\
\alpha & =\frac{\sqrt{b^{2}+a\left(q_{x}^{2}-q_{y}^{2}\right)}-b}{a}\,,\label{eq:pitch-strength-denorm}
\end{align}
where $a=q_{\hat{p}}^{2}-q_{x}^{2}$ and $b=q_{\hat{p}}q_{y}\left(1-q_{x}^{2}\right)$.

In very noisy conditions, it is possible for the periodic estimate
$\hat{\mathbf{p}}$ to have a lower coherence than the clean speech
in a band ($q_{\hat{p}}<q_{x}$). In that case, we set $r=1$ and
compute a gain attenuation term that will ensure that the stochastic
component of the enhanced speech matches the level of the clean speech
(at the expense of making the periodic component too quiet)
\begin{equation}
g^{\left(\mathrm{att}\right)}=\sqrt{\frac{1+n_{0}-q_{x}^{2}}{1+n_{0}-q_{\hat{p}}^{2}}}\,,\label{eq:extra-attenuation}
\end{equation}
where $n_{0}=0.03$ (or 15~dB) limits the maximum attenuation to
the noise-masking-tone threshold~\cite{painter2000perceptual}. For
the normal case ($q_{\hat{p}}\geq q_{x}$), then $g^{\left(\mathrm{att}\right)}=1$.

\section{DNN Model}

\label{sec:DNN-model}

\begin{figure}

\centering{\includegraphics[width=0.85\columnwidth]{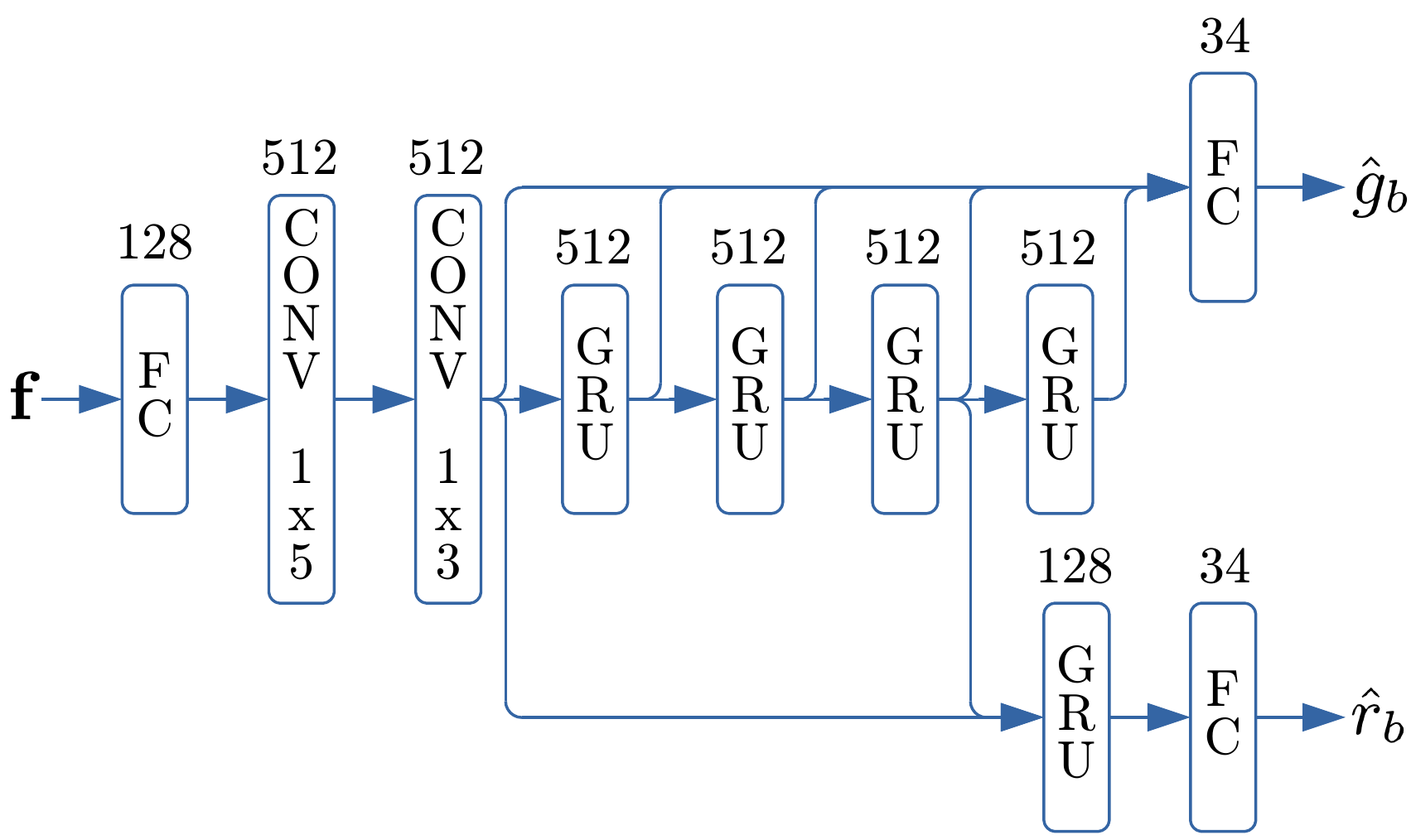}}\caption{Overview of the DNN architecture computing the 34~gains $\hat{g}_{b}$
and 34~strengths $\hat{r}_{b}$ from the 70\protect\nobreakdash-dimensional
input feature vector $\mathbf{f}$. The number of units on each layer
is indicated above the layer type.\label{fig:Overview-of-DNN}}

\end{figure}

The model uses both convolutional layers (a 1x5~layer followed by
a~1x3 layer), and GRU~\cite{cho2014properties} layers, as shown
in Fig.~\ref{fig:Overview-of-DNN}. The convolutional layers are
aligned in time such as to use up to $M$ frames into the future.
To achieve 40~ms look-ahead including the 10\nobreakdash-ms overlap,
we use $M=3$.

The input features used by the model are tied to the 34~ERB bands.
For each band, we use two features: the magnitude of the band with
look-ahead $Y_{b}\left(\ell+M\right)$ and the pitch coherence without
look-ahead $q_{y,b}\left(\ell\right)$ (the coherence estimation itself
uses the full look-ahead). In addition to those 68~band-related features,
we use the pitch period~$T\left(\ell\right)$, as well as an estimate
of the pitch correlation~\cite{vos2013} with look-ahead, for a total
of 70~input features. For each band $b$, we also have 2~outputs:
the gain $\hat{g}_{b}\left(\ell\right)$ approximates $g_{b}^{\left(\mathrm{att}\right)}\left(\ell\right)g_{b}\left(\ell\right)$
and the strength $\hat{r}_{b}\left(\ell\right)$ approximates $r_{b}\left(\ell\right)$. 

The weights of the model are forced to a $\pm\frac{1}{2}$ range and
quantized to 8-bit integers. This reduces the memory requirement (and
bandwidth), while also reducing the computational complexity of the
inference by taking advantage of vectorization.

\subsection{Training Data}

We train the model on synthetic mixtures of clean speech and noise
with SNRs ranging from -5~dB to 45 dB, with some noise-free examples
included. The clean speech data includes 120~hours of 48~kHz speech
from different public and internal databases, including more than
200~speakers and more than 20~different languages. The noise data
includes 80~hours of various noise types, also sampled at 48~kHz. 

To ensure robustness in reverberated conditions, the noisy signal
is convolved with simulated and measured room impulse responses. Inspired
by~\cite{zhao2018late}, the target includes the early reflections
so that only late reverberation is attenuated.

We improve the generalization of the model by applying a different
random second-order pole-zero filter to both the speech and the noise.
We also apply the same random spectral tilt to both signals to better
generalize across different microphone frequency responses. To achieve
bandwidth-independence, we apply a low-pass filter with a random cutoff
frequency between 3~kHz and 20~kHz. This makes it possible to use
the same model on narrowband to fullband audio.

\subsection{Loss function}

We use a different loss function for the gain and for the pitch filtering
strength. For the gain, we consider that the perceptual loudness of
a signal is proportional to its energy raised to a power $\gamma/2$,
where we use $\gamma=0.5$. For that reason, we raise the gains to
the power $\gamma$ before computing the metrics. In addition to the
squared error, we also use the fourth power to overemphasize the cost
of making large errors (e.g. completely attenuating speech):
\begin{equation}
\mathcal{L}_{g}=\sum_{b}\left(g_{b}^{\gamma}-\hat{g}_{b}^{\gamma}\right)^{2}+C_{4}\sum_{b}\left(g_{b}^{\gamma}-\hat{g}_{b}^{\gamma}\right)^{4}\,,\label{eq:gain-loss}
\end{equation}
where we use $C_{4}=10$ to balance between the $L_{2}$ and $L_{4}$
terms.

Although simple, the loss function in (\ref{eq:gain-loss}) implicitly
incorporates many of the characteristics of the improved loss function
proposed in~\cite{erdogan2018investigations}, including scale-invariance,
SNR-invariance, power-law compression, and non-linear frequency resolution.

For the pitch filtering strength, we use the same principle as for
$\mathcal{L}_{g}$ but evaluating the loudness of the noisy component
of the enhanced speech. Since the comb filter with strength $r_{b}$
attenuates the noise by a factor $\left(1-r_{b}\right)$, we use the
strength loss
\begin{equation}
\mathcal{L}_{r}=\sum_{b}\left(\left(1-r_{b}\right)^{\gamma}-\left(1-\hat{r}_{b}\right)^{\gamma}\right)^{2}\,.\label{eq:strength-loss}
\end{equation}
Since the enhancement is not overly sensitive to errors in the value
of $\hat{r}_{b}$, we do not use a fourth power term.

\section{Envelope Postfiltering}

\label{sec:post-filtering}

To further enhance the speech, we slightly deviate from the gains
$\hat{g}_{b}$ produced by the DNN. The deviation is inspired by the
formant postfilters~\cite{chen1995adaptive} often used in CELP codecs.
We intentionally de-emphasize noisier bands slightly further than
they would be in the clean signal, while overemphasizing clean bands
to compensate. This is done by computing a warped gain
\begin{equation}
\hat{g}_{b}^{(w)}=\hat{g}_{b}\sin\left(\frac{\pi}{2}\hat{g}_{b}\right)\,,\label{eq:gain-warping}
\end{equation}
which leaves $\hat{g}_{b}$ essentially unaffected for clean bands,
while squaring it (like the gain of a Wiener filter) for very noisy
bands. To avoid over-attenuating the enhanced signal as a whole, we
also apply a global gain compensation heuristic computed as
\begin{equation}
G=\sqrt{\frac{\left(1+\beta\right)\frac{E_{0}}{E_{1}}}{1+\beta\left(\frac{E_{0}}{E_{1}}\right)^{2}}}\,,\label{eq:gain-compensation}
\end{equation}
where $E_{0}$ is the total energy of the enhanced signal using the
original gain $\hat{g}_{b}$ and $E_{1}$ is the total energy when
using the warped gain $\hat{g}_{b}^{(w)}$. We use $\beta=0.02$,
which results in a maximum theoretical gain of 5.5~dB for clean bands.
Scaling the final signal for the frame by $G$ results in a perceptually
cleaner signal that is about as loud as the clean signal. The band
energy after that postfilter is given by
\begin{equation}
\hat{X}_{b}=G\hat{g}_{b}^{(w)}Y_{b}\,.\label{eq:postfilter-magnitude}
\end{equation}

When listening to the enhanced speech through loudspeakers in a room,
the impulse response of the room is added back to the signal such
that it blends with any speech coming from the room. However, when
listening through headphones, the lack of any reverberation can make
the enhanced signal sound overly \emph{dry} and unnatural. This is
addressed by enforcing a minimum decay in the energy, subject to never
exceeding the energy of the noisy speech:
\begin{equation}
\hat{X}_{b}^{(r)}\left(\ell\right)=\min\left(\max\left(\hat{X}_{b}\left(\ell\right),\delta\hat{X}_{b}^{(r)}\left(\ell-1\right)\right),\hat{Y}_{b}\left(\ell\right)\right)\,,\label{eq:postfilter-reverb}
\end{equation}
where $\delta$ is chosen to be equivalent to a reverberation time
$T_{60}=100\,\mathrm{ms}$.

After the frequency-domain enhanced speech is converted back to the
time domain, a high-pass filter is applied to the output. The filter
helps eliminating some remaining low-frequency noise and its cutoff
frequency is determined by the estimated pitch of the talker~\cite{vos2013}
to avoid attenuating the fundamental.

\section{Experiments and Results}

\label{sec:results}

We evaluate the quality of the enhanced speech with two mean opinion
score (MOS)~\cite{P.800} tests conducted using the crowdsourcing
methodology P.808~\cite{P.808}. First, we use the 48~kHz noisy
VCTK test set provided in~\cite{valentini2016investigating} to compare
PercepNet to the original RNNoise~\cite{valin2018rnnoise}, while
also conducting an ablation study. The test includes 824~samples,
rated by 8 listeners each, resulting in a 95\% confidence interval
of 0.04. We also provide PESQ-WB~\cite{P.862.2} results as a reference
for comparison with other methods like SEGAN~\cite{pascual2017segan}.
The results in Table~\ref{tab:Internal-P.808-MOS-results} not only
demonstrate a base improvement over RNNoise, but also show that both
the pitch filter and the envelope postfilter help improve the quality
of the enhanced speech. In addition, subjective testing clearly shows
the limitations of PESQ-WB when evaluating the envelope postfilter
-- even though the subjective evaluation shows a strong improvement
from the postfilter, PESQ-WB considers it a degradation. Note that
the unusually high absolute numbers in the MOS results are likely
due to the fullband samples in that test.

In the second test, the DNS challenge~\cite{reddy2020interspeech}
organizers evaluated \emph{blind} test samples processed with PercepNet
and provided us with the results in Table~\ref{tab:Challenge-official-P.808-MOS}.
The test set includes 150~synthetic samples without reverberation,
150~synthetic samples with reverberation, and 300~real recordings.
Each sample was rated by 10~listeners, leading to a 95\% confidence
interval of 0.02 for all algorithms. Since PercepNet operates at 48~kHz,
the 16\nobreakdash-kHz challenge test data was internally up-sampled
(and later down-sampled) in the STFT domain, avoiding any additional
algorithmic delay. The same model parameters were used for both the
challenge 16\nobreakdash-kHz evaluation and our own 48\nobreakdash-kHz
VCTK evaluation, demonstrating the capability to operate on speech
with different bandwidths. The quality also exceeds that of the baseline~\cite{xia2020weighted}
algorithm.

\begin{table}
\caption{P.808 MOS results based on internal testing on the VCTK test set at
48 kHz.\label{tab:Internal-P.808-MOS-results} }

\centering{%
\begin{tabular}{lcc}
\hline 
Algorithm & PESQ-WB & \textbf{MOS (P.808)}\tabularnewline
\hline 
Noisy & 1.97 & 3.40\tabularnewline
SEGAN~\cite{pascual2017segan} & 2.16 & -\tabularnewline
RNNoise (original)~\cite{valin2018rnnoise} & 2.29 & 3.70\tabularnewline
PercepNet (no pitch, no pf) & 2.64 & 3.81\tabularnewline
PercepNet (no pf) & \textbf{2.73} & 3.91\tabularnewline
PercepNet (no pitch) & 2.47 & 3.93\tabularnewline
\textbf{PercepNet} & 2.54 & \textbf{4.05}\tabularnewline
\hline 
\end{tabular}}

\end{table}

\begin{table}
\caption{Challenge official P.808 MOS results. The baseline model is provided
by the challenge organizers.\label{tab:Challenge-official-P.808-MOS}}

\centering{%
\begin{tabular}{lcccc}
\hline 
Algorithm & Synthetic & Synthetic & Real & \textbf{Overall}\tabularnewline
 & w/o reverb & w/ reverb & record & \tabularnewline
\hline 
Noisy & 3.32 & 2.78 & 2.97 & 3.01\tabularnewline
Baseline & 3.49 & 2.64 & 3.00 & 3.03\tabularnewline
\textbf{PercepNet} & \textbf{3.92} & \textbf{3.16} & \textbf{3.51} & \textbf{3.52}\tabularnewline
\hline 
\end{tabular}}
\end{table}

The algorithm complexity is mostly dictated by the neural network,
and thus the number of weights. For a frame size of 10~ms and 8M~weights,
the complexity is around 800~MMACS (one multiply-and-accumulate per
weight per frame/second). By quantizing the weights with 8~bits,
vectorization makes it possible to run the network efficiently. With
the default frame size of 10~ms, PercepNet requires 5.2\% of one
mobile x86 core (1.8~GHz Intel i7-8565U CPU) for real-time operation.
Evaluated with a frame size of 40~ms (four internal frames of 10~ms
each to improve cache efficiency), the complexity is reduced to 4.1\%
on the same CPU core with an identical output. Despite a much lower
complexity than the maximum allowed by the DNS challenge, PercepNet
ranked second in the real-time track.

Qualitatively\footnote{Audio samples available at: \href{https://www.amazon.science/interspeech-2020-deep-noise-suppression-audio-samples}{https://www.amazon.science/interspeech-2020-deep-noise-suppression-audio-samples}},
the use of ERB bands -- rather than operating directly on frequency
bins -- makes the algorithm incapable of producing musical noise (\emph{aka}
birdie artifacts) in the output. Similarly, the short window used
for analysis avoids reverb-like smearing in the time domain. Instead,
the main noticeable artifact is a certain amount of \emph{roughness}
caused by some noise remaining between pitch harmonics, especially
for loud car noise.

\section{Conclusion}

We have demonstrated an efficient speech enhancement algorithm that
focuses on the main perceptual characteristics of speech -- spectral
envelope and periodicity -- to produce high-quality fullband speech
in real time with low complexity. The proposed PercepNet model uses
a band structure to represent the spectrum, along with pitch filtering
and an additional envelope postfiltering step. Evaluation results
show significant quality improvements for both wideband and fullband
speech and demonstrate the effectiveness of both the pitch filtering
and the postfilter. We believe the results demonstrate the benefits
of modeling speech using perceptually-relevant parameters. 

\label{sec:conclusion}

\bibliographystyle{unsrt}
\bibliography{rnnoise}

\end{document}